\documentclass{ws-procs9x6}
\usepackage{overpic}
\usepackage{epsfig}
\usepackage{amssymb,amsmath}
\usepackage{ulem}
\usepackage[usenames]{color}

\newcommand{\CK}{\v Cerenkov}

\begin{document}

\title{Cosmic Ray Velocity and Electric Charge Measurements in the AMS experiment}
\vspace{-1.0cm}   
\author{Lu\'isa Arruda, \small{on behalf of the AMS-02 collaboration}}
\address{LIP/IST \\
         Av. Elias Garcia, 14, 1$^o$ andar\\
         1000-149 Lisboa, Portugal \\
         e-mail: luisa@lip.pt}
\maketitle
\abstracts{             
The Alpha Magnetic Spectrometer (AMS) is a particle physics detector designed
to measure charged cosmic ray spectra with energies up to the TeV region and
with high energy photon detection capability up to few hundred GeV. It will 
be installed on the 
International Space Station (ISS) in 2008 and will operate for more than three years.
Due to its large acceptance, the flight duration and the
state-of-art of particle identification techniques, AMS will have a remarkable
sensitivity on antimatter and dark matter searches.\\
The addition of different detector systems provide AMS
with complementary and redundant electric charge and velocity measurements. 
The velocity of singly charged particles is expected to be 
measured with a precision of 0.1$\%$ and charge separation up to iron is
attainable.
The AMS capability of measuring a large range of electric charges and accurate
velocities, will largely contribute to a better understanding of cosmic ray production,
acceleration and propagation mechanisms in the galaxy.} 
\vspace{-1.0cm}
\section{The AMS02 detector}
AMS~[\cite{bib:ams}] (Alpha Magnetic Spectrometer) is a precision spectrometer 
designed to search for cosmic antimatter, dark matter and to study the
relative abundance of elements and isotopic composition of the primary cosmic
rays with an energy up to $\sim$1\,TeV. 
It will be installed in the International Space Station (ISS), in 2008, where
it will operate for more than three years.

\begin{figure}[htb]
\begin{center}
\vspace{-0.5cm}
\epsfig{file=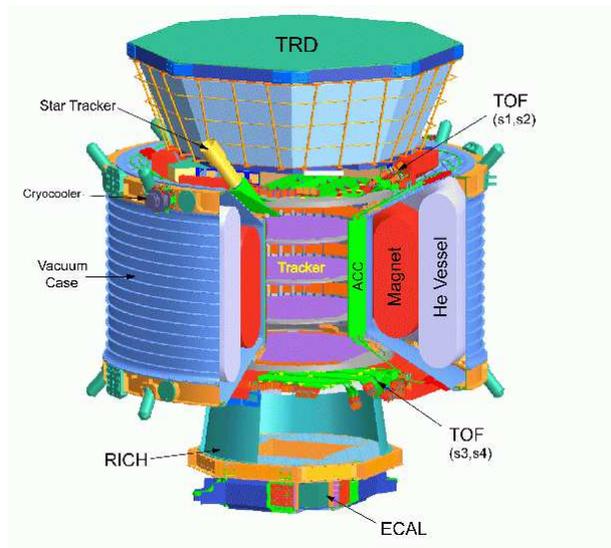,width=0.71\linewidth,clip=,bbllx=0,bblly=0,bburx=612,bbury=550}   
\caption{A whole view of the AMS Spectrometer.}
\end{center}  
\label{fig:ams}
\end{figure}

The spectrometer  will be able to measure
the rigidity ($R\equiv pc/ |Z| e$), the charge ($Z$),
the velocity ($\beta$) and the energy ($E$) of cosmic rays within a 
geometrical acceptance of $ \sim$0.5\,m$^2$.sr.
Figure \ref{fig:ams} shows a schematic view of the AMS spectrometer.
At both ends of the AMS spectrometer exist the 
Transition Radiation Detector
(TRD) (top) and the Electromagnetic Calorimeter (ECAL) (bottom). Both will provide AMS with capability to
discriminate between leptons and hadrons. 
Additionally the calorimeter will trigger and detect photons.
The TRD will be followed by the first of the four Time-of-Flight (TOF) system 
scintillator planes. 
The TOF system~[\cite{bib:ams02-tof}] is composed of four roughly circular planes of 12\,cm wide
scintillator paddles, one pair of planes above the magnet, the upper TOF,
and one pair bellow, the lower TOF. There will be a total of 34 paddles. The
TOF will provide a fast trigger within 200\,ns, charge and velocity measurements for charged particles, 
as well as information on their direction of incidence. The TOF operation at
regions with very intense magnetic fields forces the use of shielded fine mesh
phototubes and the optimization of the light guides geometry, with some of
them twisted and bent. Moreover the system guarantees redundancy, with two photomultipliers
on each end of the paddles and double redundant electronics. A time
resolution of 130\,ps for protons is expected~[\cite{bib:ams02-note}].\\
The tracking system will be surrounded by veto counters and
embedded in a magnetic field of about 0.9\,Tesla produced by a 
superconducting magnet.
It will consist on a Silicon Tracker~[\cite{bib:ams02-tracker}] 
made of 8 layers of double sided silicon 
sensors with a total area of $\sim$6.7\,m$^2$. There will be a total of
$\sim$2500 silicon sensors arranged on 192 ladders. The position of the charged particles crossing
the tracker layers is measured with a precision of $\sim$10\,$\mu$m along
the bending plane and $\sim$30\,$\mu$m
on the transverse direction. 
With a bending power (BL$^2$) of around 0.9\,T.m$^2$, particles rigidity is measured    
with an accuracy better than 2\% up to 20\,GV and the maximal detectable rigidity is
around 1-2\,TV. Electric charge is also measured from energy deposition up
to Z$\sim$26.\\ 
The Ring Imaging \CK\ Detector (RICH)~[\cite{bib:buenerd}]
will be located right after the last TOF plane and before the electromagnetic
calorimeter. It is a
proximity focusing device with a dual radiator configuration on
the top made of a low refractive index aerogel 1.050, 2.7\,cm thick and a central square
of sodium fluoride (NaF), 0.5\,cm thick. Its detection matrix is composed of
680 photomultipliers and light guides and a high reflectivity conical mirror
surrounds the whole set. RICH was designed to measure the velocity
of singly charged particles with a resolution $\Delta\beta/\beta$ of 0.1$\%$,
to extend the charge separation up to iron, to contribute to $e/p$ separation 
and to albedo rejection.\\
Particle identification on AMS-02 relies on a very precise determination 
of the magnetic rigidity, energy, velocity and electric charge.
Velocity of low energy particles (up to $\sim 1.5$ GeV) is measured by TOF detector while for
kinetic energies above the radiator thresholds (0.5\,GeV for sodium fluoride and 2\,GeV for aerogel) 
the RICH will provide very accurate measurements; a target resolution of $\sim$1$\%$ and $\sim$0.1$\%$ 
for singly charged particles is expected, respectively  
for sodium fluoride and aerogel radiators.   
The electric charge is measured by the silicon tracker and TOF detectors through $dE/dx$ samplings and
on the RICH through the \CK\ ring signal integration.
Charge identification, at least, up to the iron element is expected. 
\vspace{-1.1cm} 
\section{Velocity measurement}
\begin{figure}[htb]
\begin{center}
\hspace{-.3cm}
\begin{tabular}{cc}
\scalebox{0.9}{%
\scalebox{0.3}{%
\includegraphics[bb=5 -25 519 487]{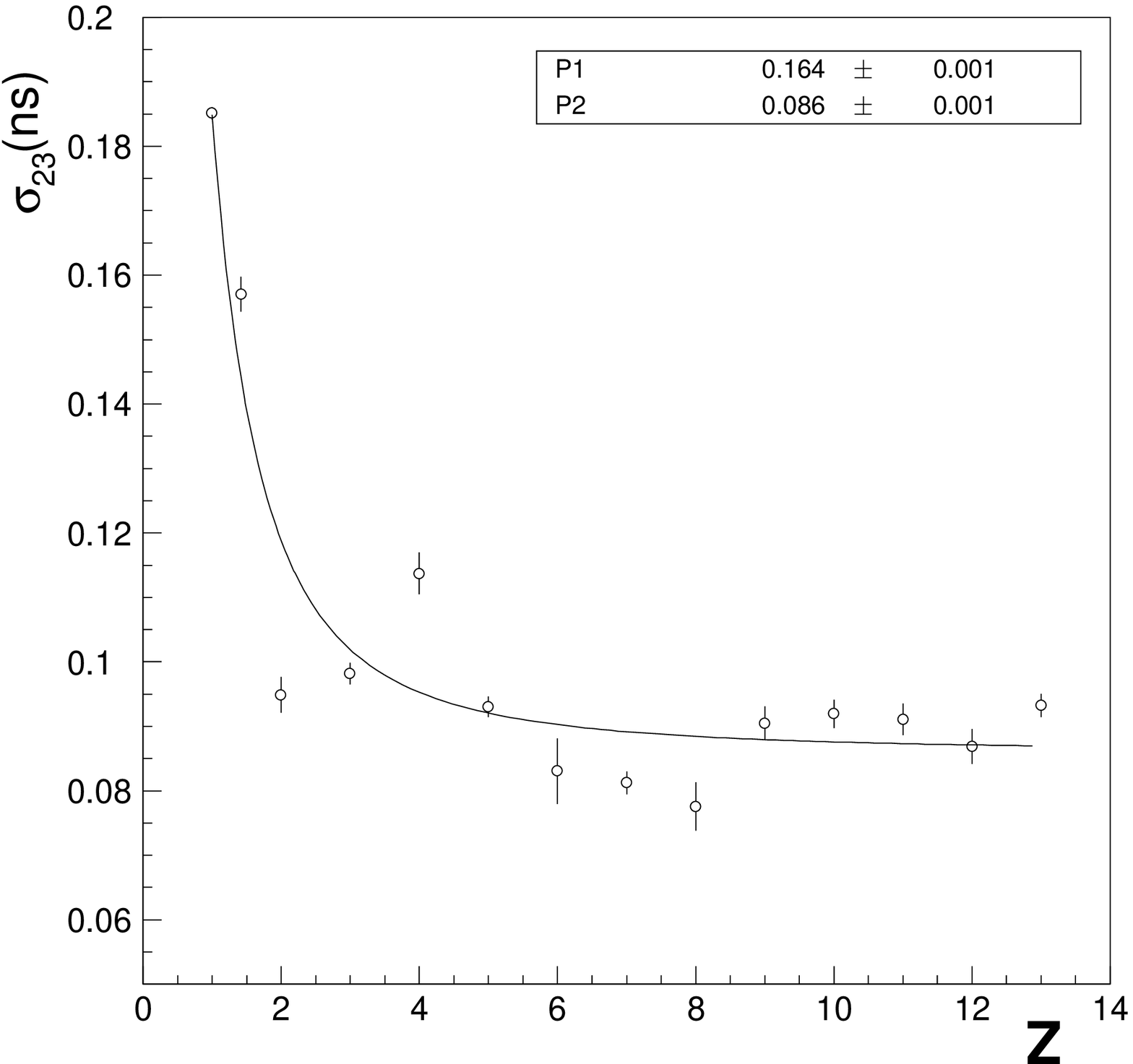}
}
\hspace{-.2cm} 
\scalebox{0.36}{%
\begin{overpic}{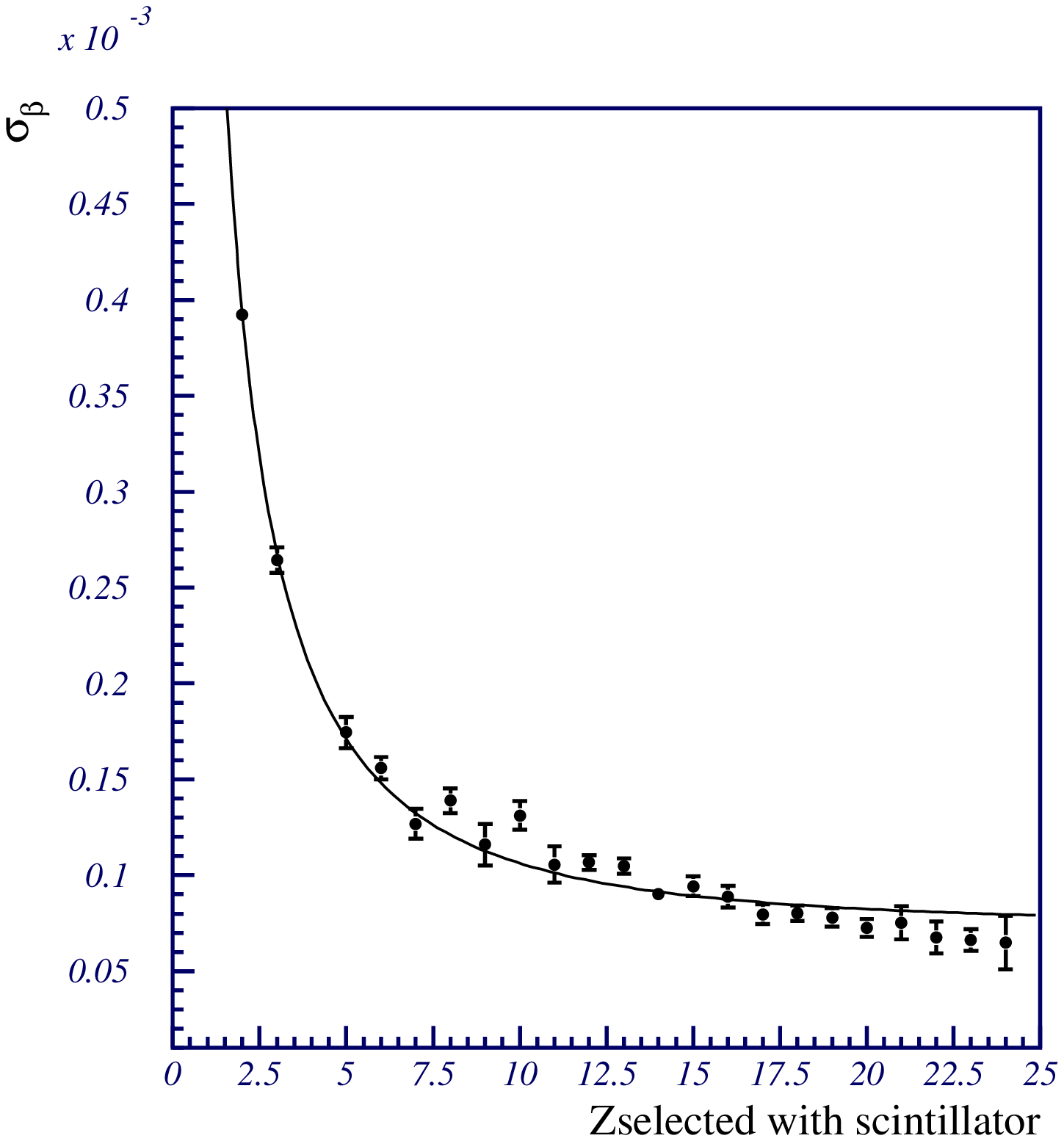} 
\put(40,75){\huge{$\sqrt {({\frac{A}{Z}})^{2}+B^{2}}$}}
\put(40,65){\huge{$A=(7.77 \pm 0.05)\times10^{-4}$}}
\put(40,58){\huge{$B=(7.3 \pm 0.2)\times10^{-5}$}}
\put(40,51){\huge{$\chi^{2}=55.2/20$}}
\end{overpic}
}
}
\end{tabular} 
\vspace{-0.2cm}
\caption{Time of flight resolution for a set of two scintillators and different charged nuclei (left), 
and evolution of the $\beta$ resolution with the charge obtained for a RICH
prototype with an aerogel 1.03 radiator, 3\,cm thick.
Both results are from 
nuclei fragments of an indium beam of 158\,GeV/c/nuc taken at CERN in October 2003.
\label{fig:thc}}
\end{center}
\end{figure}

TOF measures the crossing time between two scintillator planes and extracts
the velocity through $\beta$=$\Delta$L/$\Delta$t. The time of flight
resolution for two scintillators, tested in a test beam at CERN in October 2003 with 
fragments of an indium beam of 158\,GeV/c/nuc, is shown in figure \ref{fig:thc} as 
function of the particle charge. 
One of the tested scintillators had bent and twisted 
light guides (C2) while the other one had bent light guides (C3). 
A time resolution of 180\,ps was estimated for this conservative configuration. 
However, as the measurement in AMS-02 will be
done with four independent measurements, the time resolution which can be
inferred is of the order of 130\,ps for a minimum ionizing particle.\\
In the RICH detector the velocity of the particle, $\beta$, 
is straightforwardly
derived from the \CK\ angle reconstruction ($\cos\theta_c=\frac{1}{\beta~n}$). 
Two reconstruction methods were developed: 
a geometrical method based on a hit-by-hit reconstruction and a method 
 based on a likelihood fit to the pattern of 
the detected photons.
The best value of $\theta_c$ will result in the former case from the average
of the
hit velocities after hit clusterization and in the latter 
from the maximization
of a Likelihood function $L(\theta_c)$ given by, 
\vspace{-0.25cm}
\begin{equation}
L(\theta_c) = \prod_{i=1}^{nhits} p_i^{n_i}  \left[ r_i(\theta_c) \right].
\label{eq:likelihood}
\vspace{-0.3cm}
\end{equation}
\noindent where the probability
of a hit belonging to a \CK\ ring of angle $\theta_c$ ($p_i$) is function
of the closest distance of the hit to the \CK\ pattern ($r_i$). 
In both
methods the hits position is weighted by the detected signal $n_i$.
For a more complete description of the method see [\cite{bib:NIM}].
The resolution expected for $\beta\sim$1 singly charged particles 
crossing the aerogel
radiator is around $4\,mrad$ while for those crossing 
the NaF radiator is around $8\,mrad$. 
The accuracy of the velocity determination improves with the charge.
Figure \ref{fig:thc} (right plot) shows the evolution of the velocity resolution
for indium beam fragments of $158 ~GeV/c/n$ detected with a RICH prototype 
corresponding to $1/6$ of the final RICH detector.
The radiator plane was placed perpendicular to the beam direction. 
The photon expansion length between the
radiator and the detection matrix was adjusted to 42.3\,cm, 
allowing fully contained \CK\ rings.
\vspace{-1.0cm}
\vspace{-0.8cm}
\section{Charge measurements}
\begin{figure}[htb]
\begin{center}
\vspace{-0.5cm}
\scalebox{0.9}{%
\begin{tabular}{cc}
\hspace{-0.9cm}
\scalebox{0.3}{%
\begin{overpic}{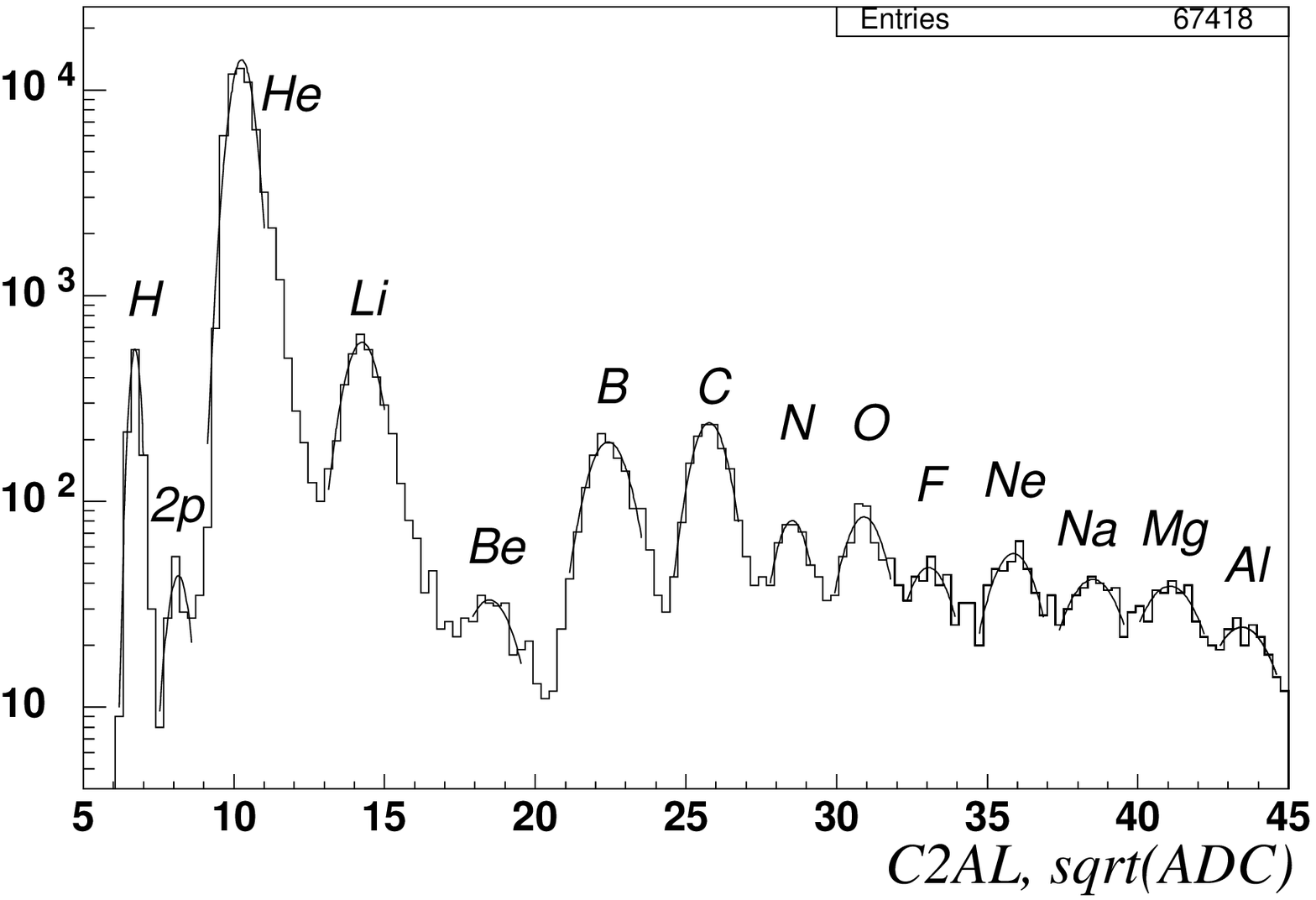} 
\put(88,60){\textbf{\Huge{a}}}
\end{overpic} 
}
\hspace{0.3cm}
\scalebox{0.29}{%
\begin{overpic}[bb=14 30 582 379]{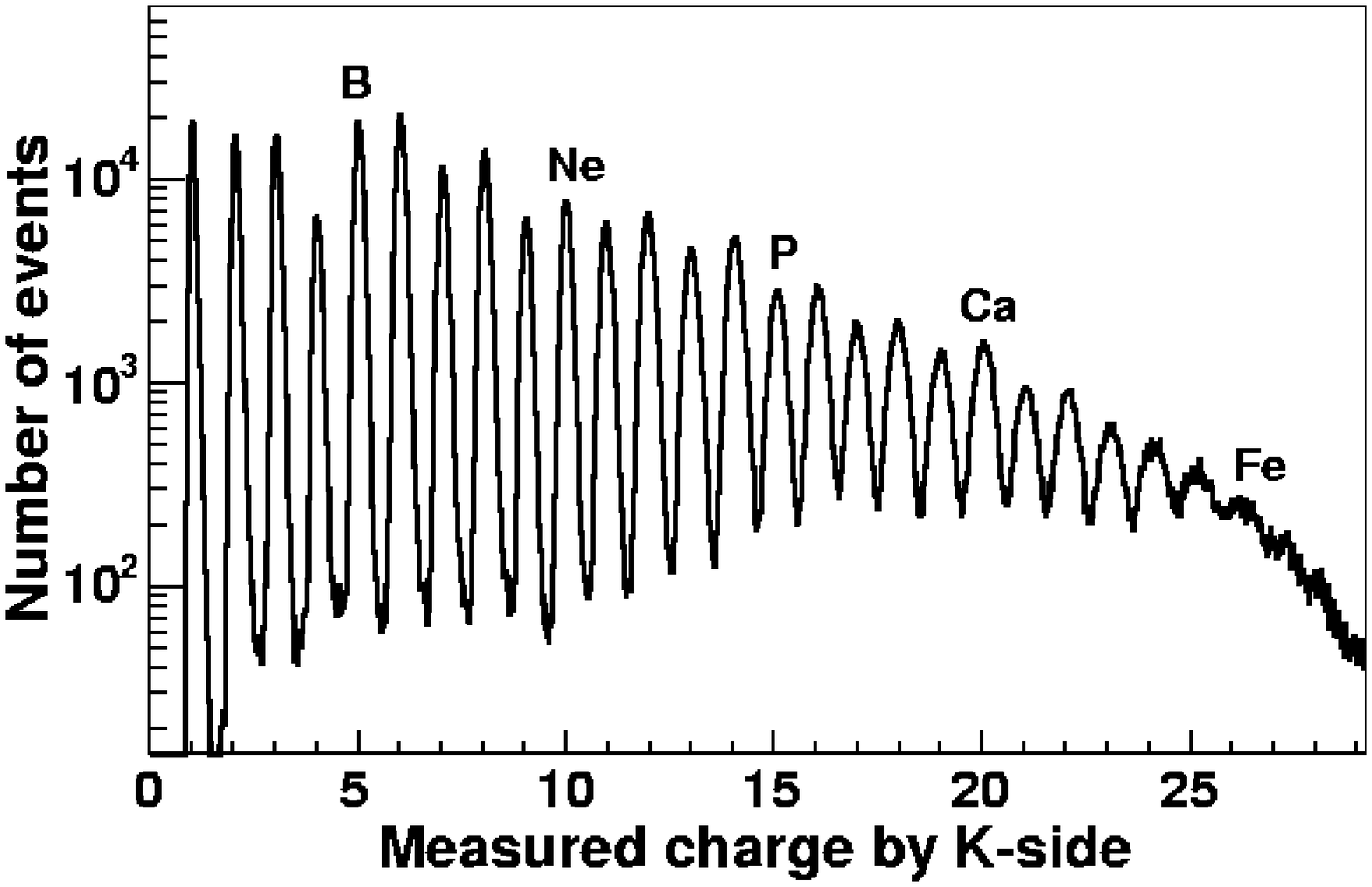}
\put(89,54){\textbf{\Huge{b}}}
\end{overpic}  
}
\\
\hspace{0.2cm}
\scalebox{0.3}{%
\begin{overpic}{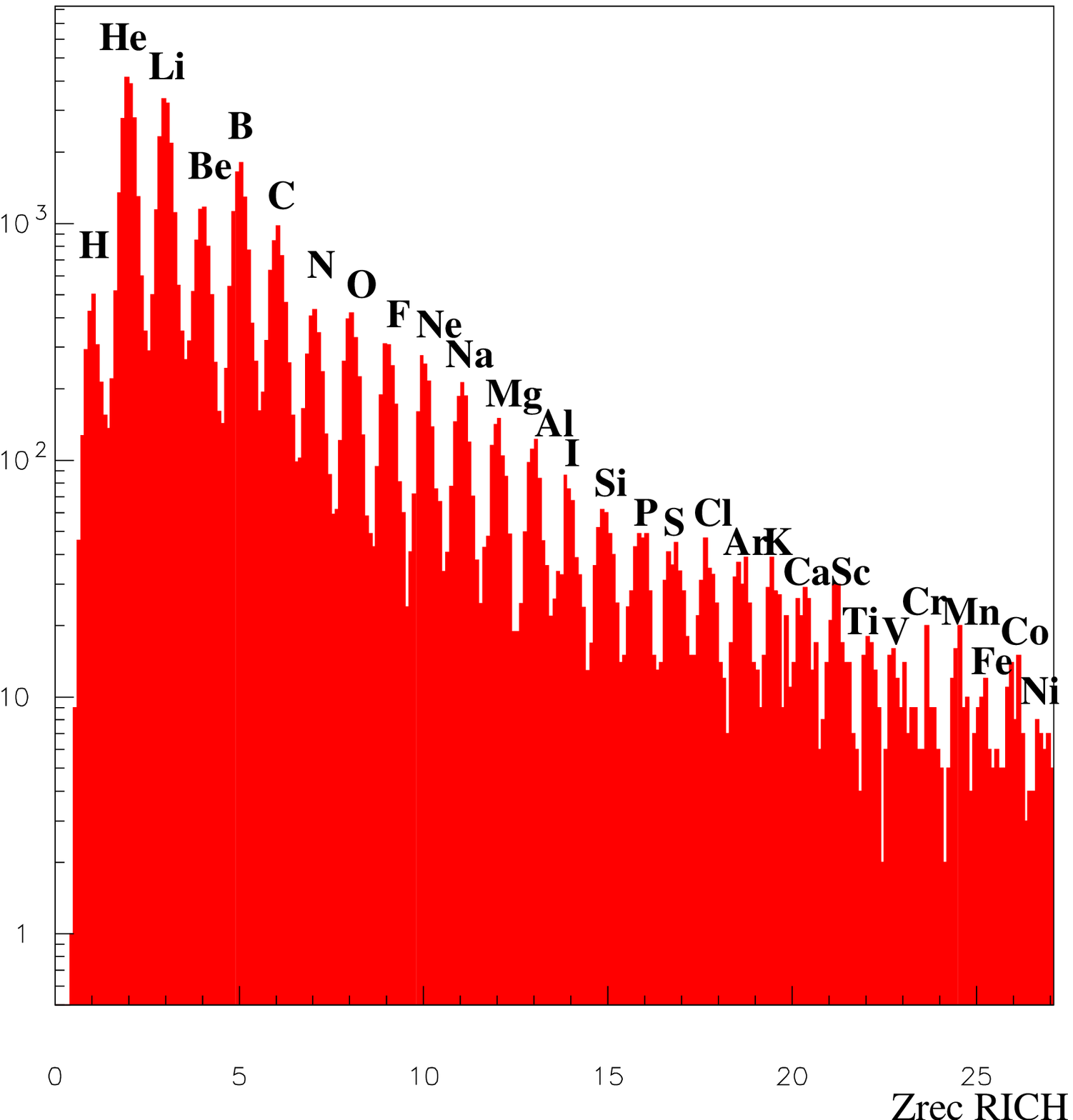}
\put(85,90){\textbf{\Huge{c}}}
\end{overpic}
}             
\hspace{0.1cm}
\scalebox{0.35}{%
\begin{overpic}[bb=0 20 567 464]{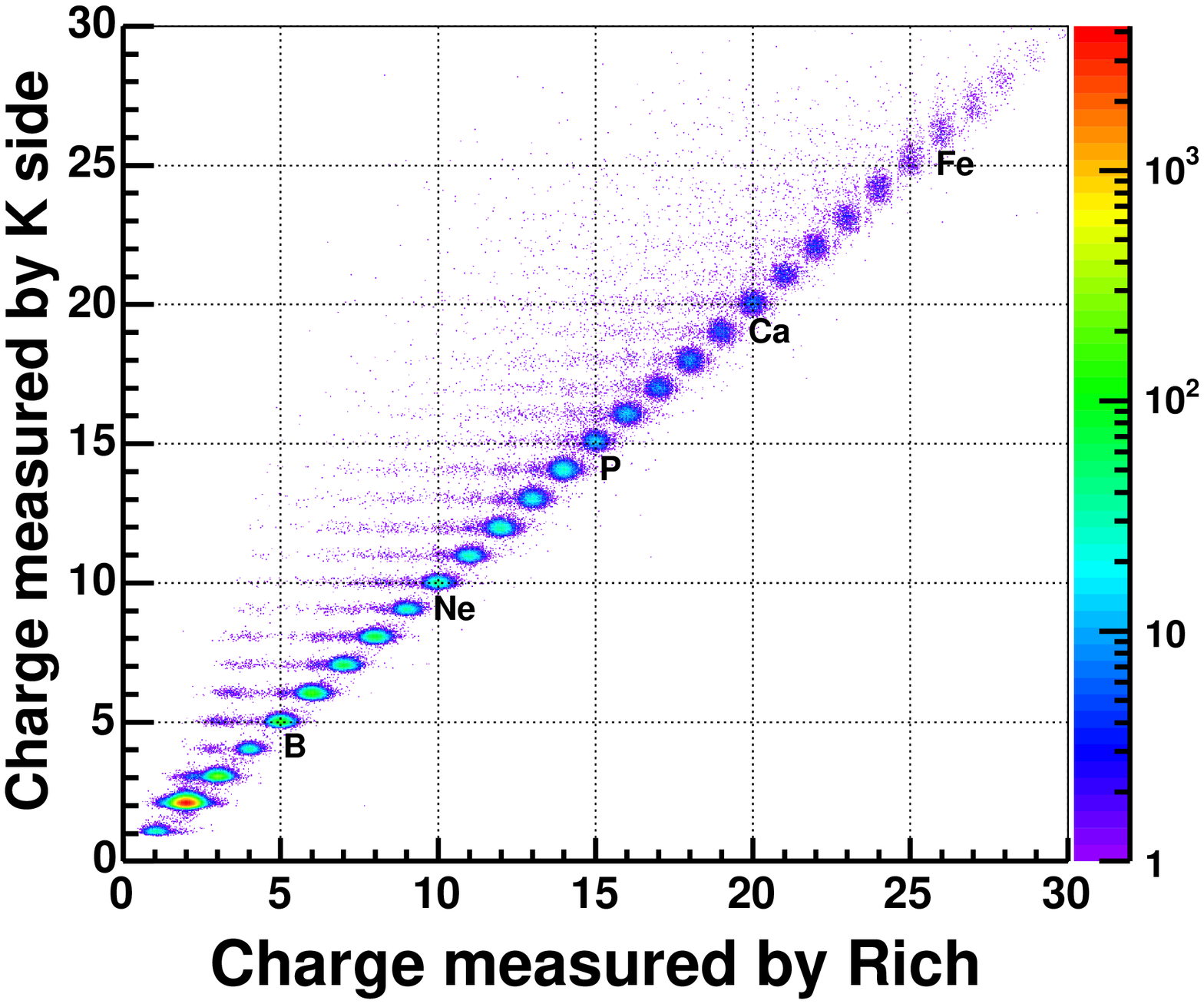}
\put(75,70){\textbf{\Huge{d}}}
\end{overpic} 
}
\end{tabular} 
}
\vspace{-0.2cm}
\caption{
  Charge measurements with TOF, Tracker and RICH prototypes tested
   at CERN in October 2003, with fragments of an indium beam of 
158\,GeV/c/nuc. (a) The square root of the integrated charge
  measured with TOF scintillator's anode show different peaks
  corresponding to charges up to aluminium. (b) Combined Z
  measurements for 4 or more ladders on the tracker K side. 
  (c) Charge peaks distribution measured with the  RICH prototype having an n=1.05 aerogel radiator, 
2.5\,cm thick. 
  (d) Comparison of the charge measurements made by the tracker and by the RICH. \label{fig:chg}}
\end{center}
\vspace{-0.5cm}
\end{figure}

As it was said previously, TOF and tracker measure the charge (Z) through $dE/dx$
samplings. Figure \ref{fig:chg} a) shows the charge measurement from the
anode signal of one of the TOF counters (C2)
tested in ion beam at CERN in 2003, in principle the most unfavourable
one. 
Charge separation up to the aluminium is visible.
Figure \ref{fig:chg} b) shows the combined measurements for 4 or more ladders on the tracker K
side. The ion species can be distinguished up to Z=26. 

\noindent RICH charge measurement is based on the fact that the number of \CK\ photons produced in the radiator 
depends on the particle charge through
$N_{\gamma} \propto Z^2 L \sin^2\theta_c$
where $L$ is the radiator thickness.
Once the total number of photoelectrons ($N_{p.e}$) associated to a \CK\ ring is computed one has to 
correct it by  the photon ring overall efficiency in order to derive the charge.   
The uncertainty on charge determination results from two distinct contributions. 
One of statistical nature independent of the nuclei charge and depending essentially   
on the amount of \CK\ signal detected for singly charged particles  ($N_{p.e} \sim 10$).
Another one of systematic nature scaling with the charge and coming essentially from 
non-uniformities on the radiator plane and photon detection efficiency.
The RICH goal of a good charge separation in a wide range of nuclei charges implies 
a good mapping and monitoring of the potential non-uniformities present on the detector.

Charge peaks reconstructed with the RICH prototype for data taken at CERN during October 2003, 
are shown in figure \ref{fig:chg} c).
The RICH configuration included an aerogel radiator of n=1.05 and 2.5\,cm thick.
A charge resolution for helium events  slightly  better than $\Delta Z \sim 0.2$ 
was observed together with a systematic uncertainty of $1\%$. A clear charge separation up to Z=28 was achieved.
For a more complete description of the charge reconstruction method see
[\cite{bib:NIM}]. Finally  figure \ref{fig:chg} d) presents the comparison of
charge measured by tracker and RICH. Ions can be distinguished and identified
up to Z=26, an excellent correlation is obtained.
\vspace{-0.1cm}
\section{Acknowledgements}
I would like to express my acknowledgements to the AMS collaboration for giving me
the opportunity to attend the Lake Louise Winter Institute 2005. I also
express my gratitude to the Funda\c{c}\~ao da Ci\^encia e Tecnologia for the
financial support to participate in the meeting.
\vspace{-0.1cm}
\section{Conclusions}
AMS is a spectrometer designed for antimatter and dark matter searches and for measuring
relative abundances of nuclei and isotopes.The velocity of singly charged particles is expected to be 
measured with a precision of 0.1$\%$ and charge separation up to iron is
attainable. Real data analysis was done with data collected with prototypes
of TOF, tracker and RICH in test beams at CERN, in October 2002 and 2003\\

\end{document}